# Nonlinear force dependence on optically bound micro-particle arrays in the evanescent fields of fundamental and higher order microfibre modes

Aili Maimaiti[1,2], Daniela Holzmann[3], Viet Giang Truong[1], Helmut Ritsch[3] & Síle Nic Chormaic[1*]

[1]Light-Matter Interactions Unit, Okinawa Institute of Science and Technology Graduate University, Onna, Okinawa 904-0495, Japan
[2]Physics Department, University College Cork, Cork, Ireland
[3]Institute for Theoretical Physics, University of Innsbruck, Technikerstrasse 25, A-6020, Innsbruck, Austria
*Correspondence to: sile.nicchormaic@oist.jp

**Abstract**

Particles trapped in the evanescent field of an ultrathin optical fibre interact over very long distances via multiple scattering of the fibre-guided fields. In ultrathin fibres that support higher order modes, these interactions are stronger and exhibit qualitatively new behaviour due to the coupling of different fibre modes, which have different propagation wave-vectors, by the particles. Here, we study one dimensional longitudinal optical binding interactions of chains of 3 µm polystyrene spheres under the influence of the evanescent fields of a two-mode microfibre. The observation of long-range interactions, self-ordering and speed variation of particle chains reveals strong optical binding effects between the particles that can be modelled well by a tritter scattering-matrix approach. The optical forces, optical binding interactions and the velocity of bounded particle chains are calculated using this method. Results show good agreement with finite element numerical simulations. Experimental data and theoretical analysis show that higher order modes in a microfibre offer a promising method to not only obtain stable, multiple particle trapping or faster particle propulsion speeds, but that they also allow for better control over each individual trapped object in particle ensembles near the microfibre surface.

**Introduction**

Optical trapping with a tightly focussed laser beam was first reported by Ashkin *et al*.[1]. Following this early work, optical tweezers have been widely used and further developed to provide stable trapping and manipulation of small objects[2]. Shortly after Ashkin *et al*.'s pioneering work, the self-ordered distribution of particles in the maxima of an optical lattice formed by the interference of up to five beams was demonstrated by Burns *et al*[3]. This multi-particle self-arrangement was attributed to the electromagnetic field redistribution caused by each particle due to the presence of neighbouring particles. The observed effect was termed "optical binding". More than a decade later, the observation of long-range, one dimensional (1D) longitudinally optically bound chains of microparticles in the field of two weakly-focussed counter propagating Gaussian[4] and non-diffracting Bessel beams[5] was reported. In those works, the net scattering force from each beam was cancelled out and the coherently scattered light from each trapped particle interfered to create attractive or repulsive forces between the particles. The balance between these forces of the dispersed particles caused them to re-arrange their positions with preferential interparticle spacings that were roughly equal to several times the particle diameter.

Following this, there have been many theoretical and experimental studies explaining this scattered field interaction with regard to dielectric objects[6-11], cells[12] in free space beams[6–8], as well as in the evanescent fields of prisms[9–10] and waveguides[11-13].

Unlike a 3D optical binding geometry - where the interaction between the trapped objects rapidly decays with distance - long-range, strong binding interactions can be realised if scattered polarisable particles are spatially confined in a 1D geometry[14]. In certain circumstances it could provide ultrastrong and self-consistent traps for small particles, or even atomic ensembles, if the light fields are optically confined to resonators[15] or waveguides[16-17]. Along with the use of prisms to generate evanescent fields, optical micro/nanofibres (MNFs) are attractive alternatives[18-20]. The tightly confined optical fields at the waist regions of MNFs make them distinctly valuable for a wide range of applications such as propelling dielectric and biological particles in liquid dispersion[21-23], particle sorting[24], and cold atom characterisation, trapping and detection[25–29].

Recently, Frawley *et al.* reported on optical binding of silica spheres on the surface of a nanofibre[30]. The discussion focussed on how a fundamental fibre mode (FM) can interact with surrounding objects, causing mutual interactions between them. There have been several theoretical proposals on using the first group of higher order fibre modes (HOMs) for particle trapping and detection, particularly in relation to cold atoms[31-34]. Here, HOMs imply the coexistence of true $TE_{01}$, $TM_{01}$, and $HE_{21o,e}$ modes of a fibre, also referred to as the first order linearly polarized (*$LP_{11}$*) mode family. Preliminary experimental studies on the interaction of nanofibre HOMs with cold atoms[35] and particle manipulation[36] have been published. These works illustrate the advantages of HOM-supporting MNFs, such as achievable higher field amplitudes, larger field extensions from the fibre surface, and a larger fibre taper cut-off diameter compared to that needed for FM propagation. Most importantly, the 3D geometries that can be obtained from the interference of co-propagating HOMs and FMs could facilitate studies in the retrieval and storage of orbital angular momentum of light in atomic ensembles near an MNF surface.

While previously published work has shown that the speed of single particles for HOM compared to FM propagation is increased[36], our recent study on two-particle binding in a HOM[37] field manifested new phenomena - the speed of two coupled particles and their inter-particle distances are clearly different from those of the fundamental mode case.

In this work, we study the dynamics of longitudinal, self-ordered structures of dielectric microparticles under the influence of the FM and HOMs of a 2 µm tapered fibre, both theoretically and experimentally. The first group of HOMs, which corresponds to the *$LP_{11}$* mode family, was generated by launching a first-order Laguerre-Gaussian beam (*$LG_{01}$*) into a suitable fibre. The

$LG_{01}$ beam was formed using a spatial light modulator (SLM). With the assistance of a custom-built optical tweezer it was possible to trap and move a fixed number of particles to the fibre. Particle speeds and inter-particle distances were compared for each chain. Additionally, the experimental data is supported by analytical and numerical analyses based on an intuitive three-mode scattering-matrix model for large particle ensembles on the microfibre surface. The optical binding forces on the particles for the HOMs and the FM are also verified using the full Maxwell stress tensor method.

**Theoretical analysis**

**Scattering-matrix approach and forces acting on multi-particle trapping**

In order to calculate the optical forces acting on the particles, one first needs to calculate the *E*-field surrounding them by solving the scattering problem and then use these fields to obtain the forces. For this calculation, we first develop a 2D numerical model based on the finite element method (FEM) to calculate the field distribution and the forces exerted on the particles. This method of numerical analysis is similar to that reported in our previous work[30,36]. Parameters used in this numerical simulation are: the propagating wavelength, $\lambda$ =1064 nm, the refractive index of the fibre, $n_{fibre}$ = 1.456, of water, $n_{water}$ = 1.33, and of the beads, $n_{bead}$ = 1.57. In order to be consistent with the experimental data, in all simulations the propagating power at the fibre waist was assumed to be 30 mW. Although this FEM model can be used to investigate the interaction of the scattering objects and the guided surface modes very accurately, it requires high performance computing resources even for symmetrical configurations. Therefore, we use this method to calculate the forces only for the case of a single particle and two particles. These numerical results are used to precisely determine the mode propagation loss and mode coupling strengths for the case of several trapped particles on a fibre surface.

To obtain the scattered field solution for more than two particles trapped in the evanescent fields of a microfibre, a tritter scattering-matrix approach is proposed[38]. Since the chosen trapped particles are larger than the excitation wavelength, the forward scattered light from the particles is much stronger than the backward scattering. In this alternative force calculation method, we make the simplifying assumption that the backward scattered light from particles is negligible. The physical model consists of *N* dielectric particles that are longitudinally bounded along the fibre surface as specified in Figure 1.

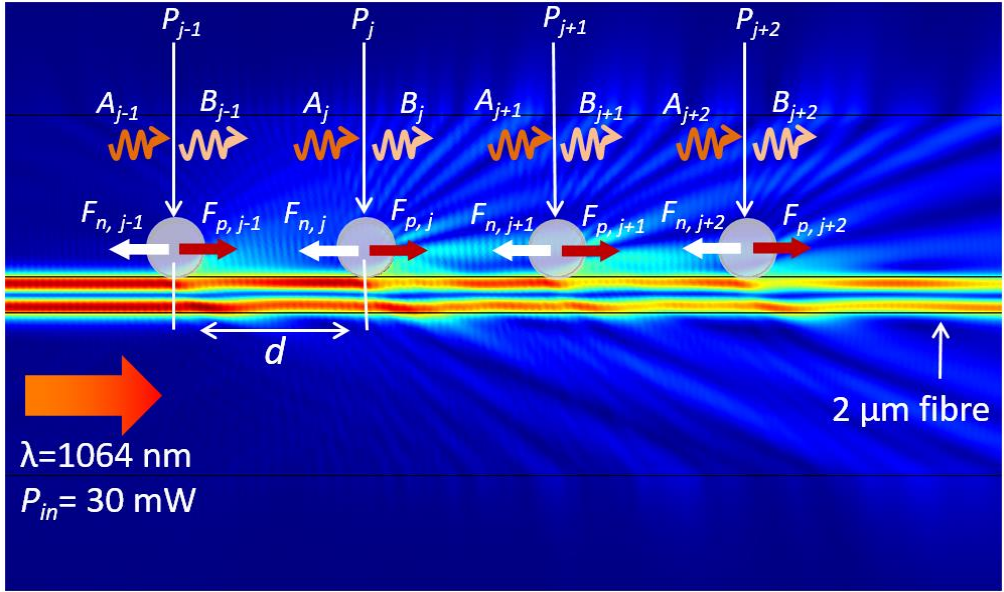

Figure 1. 1D array of *N* particles scattering light under the influence of the HOM evanescent fields of a 2 μm fibre. Laser light is coupled into the microfibre from the left at a wavelength of 1064 nm. The power at the waist is $P_{in}$ = 30 mW, $F_n$ (white arrow) and $F_p$ (red arrow) indicate the attractive and repulsive binding forces; $A_j$ (orange sinusoidal arrow) and $B_j$ (pink sinusoidal arrow) are the amplitudes of the incoming and outgoing light fields from the particles; *d* is the relative distance between the particles.

Figure 1 shows an example of how the laser light incident from the left is coupled into the microfibre and represented by the electromagnetic field, $E(x)$, at a chosen wavelength, $\lambda$ = 1064 nm. Here, the particles act as beam splitters with the input incident beam split into three beams. The sinusoidal orange ($A_j$) and pink ($B_j$) arrows represent the total incoming and outgoing light fields, respectively. The white and red arrows indicate the attractive ($F_{n,j}$) and repulsive ($F_{b,j}$) optical binding forces, respectively. For a particle at position, $x_j$, along the fibre axis, the *E*-field can be written as:

$$E(x) = A_j \exp\left(i \sum_{q=a,b,c} k_q(x-x_j)\right) + B_j \exp\left(-i \sum_{q=a,b,c} k_q(x-x_j)\right), \tag{1}$$

for *j* = 1 to *N* and $k_{tot} = \{k_a, k_b, k_c\}$ represents the complex wave-vector of three forward scattered modes propagating both inside and outside the fibre surface. When a fibre mode evanescently interacts with particles, the scattered light can be described by a wave-vector, $k_b = k_a/cos(\alpha_0)$, where $k_a$ is the wave-vector length of the incident field in the medium surrounding the particles and $\alpha_0$ is the polar angle between the incident and the scattered light. The parameter $k_c$ is the complex wave-vector component attributed to the effective propagation lossy modes which penetrate into the surrounding host. As has been proposed by Schnabel *et al.*[38], for general three-port beam-splitters, if the phases are chosen such that the scattering-matrix, **M**, is symmetric, the complex component, $\mathbf{B_j} = \mathbf{M} \times \mathbf{A_j}$, can be written as

$$\begin{pmatrix} B_{ja} \\ B_{jb} \\ B_{jc} \end{pmatrix} = \begin{pmatrix} \eta_1 e^{i\phi_0} & \eta_4 e^{i\phi_1} & \eta_5 e^{i\phi_3} \\ \eta_4 e^{i\phi_1} & \eta_2 e^{i\phi_0} & \eta_6 e^{i\phi_2} \\ \eta_5 e^{i\phi_3} & \eta_6 e^{i\phi_2} & \eta_3 e^{i\phi_0} \end{pmatrix} \begin{pmatrix} A_{ja} \\ A_{jb} \\ A_{jc} \end{pmatrix}, \qquad (2)$$

where $0 < \eta_i < 1$ describes the amplitude and $e^{i\phi_i}$ represents the phase coupling. $A_{ja}$, $A_{jb}$ and $A_{jc}$ are the incoming field components of the scattered beams on particle $j$ that allow us to identify the complex outgoing fields $B_{ja}$, $B_{jb}$ and $B_{jc}$. The outgoing field $B_j = \{B_{ja}, B_{jb}, B_{jc}\}$ is then, in turn, used as an input to the neighbouring particle 'beam splitter' via:

$$\begin{pmatrix} A_{ja+1} \\ A_{jb+1} \\ A_{jc+1} \end{pmatrix} = \begin{pmatrix} e^{ik_a(x_{j+1}-x_j)} & 0 & 0 \\ 0 & e^{ik_b(x_{j+1}-x_j)} & 0 \\ 0 & 0 & e^{ik_c(x_{j+1}-x_j)} \end{pmatrix} \begin{pmatrix} B_{ja} \\ B_{jb} \\ B_{jc} \end{pmatrix}. \qquad (3)$$

To simplify our calculation, we set the initial transmitted light phase, $\phi_0$, to be zero. In an effective 1D configuration of bounded microparticles on a microfibre surface, the amplitudes $\eta_4$, $\eta_5$ and $\eta_6$ can be assigned to the particles' scattered mode coupling strengths between the fundamental and the higher order, the higher order and the free-space, and the fundamental and the free-space modes, respectively. Here we assume that $z_l$ is the effective background scattering that represents the coupling to the lossy modes. For the unitary condition, $\mathbf{M^{-1}M} = 1$, the transmitted amplitudes can be expressed as:

$$\begin{aligned} \eta_1^2 &= 1 - \eta_4^2 - \eta_5^2 - z_l^2, \\ \eta_2^2 &= 1 - \eta_4^2 - \eta_6^2 - z_l^2, \\ \eta_3^2 &= 1 - \eta_5^2 - \eta_6^2 - z_l^2. \end{aligned} \qquad (4)$$

The phases of the scattering matrix, **M**, can be written as:

$$\begin{aligned} \phi_1 &= \tfrac{1}{2} \arccos\left( \tfrac{\eta_1^2 \eta_4^2 + \eta_2^2 \eta_4^2 - \eta_5^2 \eta_6^2}{2 \eta_4^2 \eta_1 \eta_2} \right) - \tfrac{\pi}{2}, \\ \phi_2 &= \tfrac{1}{2} \arccos\left( \tfrac{\eta_4^2 \eta_6^2 - \eta_1^2 \eta_5^2 - \eta_3^2 \eta_5^2}{2 \eta_5^2 \eta_1 \eta_3} \right), \\ \phi_3 &= \tfrac{1}{2} \arccos\left( \tfrac{\eta_2^2 \eta_6^2 + \eta_3^2 \eta_6^2 - \eta_4^2 \eta_5^2}{2 \eta_6^2 \eta_2 \eta_3} \right) + \tfrac{\pi}{2}. \end{aligned} \qquad (5)$$

The scattering force acting on the particle, $P_j$, along the fibre axis can finally be calculated from the Maxwell stress tensor and is given by:

$$F_s = \tfrac{1}{2} \varepsilon_0 \varepsilon_r \left( |A_j|^2 - |B_j|^2 \right), \qquad (6)$$

where $\varepsilon_0$ and $\varepsilon_r$ are the vacuum and the relative surrounding medium permittivity, respectively[6]. In the following section, we investigate the optical binding forces on several particles trapped along the fibre surface. Out of these, the particle-field dynamics, numerous stable configurations of particles and the influence of these effects on speed variations of particle chains are carefully discussed.

## Results

### Optical binding forces between trapped particles within a chain

In Figure 2 we present the optical binding forces and their corresponding potentials as a function of the interparticle separation, $d$, for both the FM and HOM cases. The position of the first particle, $P_1$, is fixed with respect to the fibre and the scattering forces are calculated while varying the position of the next particles, $P_j$, along the fibre axis. The longitudinal optical binding forces can be extracted by subtracting the scattering forces on the first particle from those on the next neighbouring particles. The solid and dashed lines in Figure 2 (a & d) represent the binding forces which are calculated for a two-particle model based on the FEM and the scattering-matrix methods. These figures show that the optical binding forces modulate around zero as a function of the distance between two particles. Positive and negative values represent repulsive and attractive forces on the trapped particles. A stationary particle order can be achieved if the repulsive and attractive forces vanish. When using the FEM method, there exists a short periodic oscillation in the binding forces, where the peak-to-peak distance is close to $\lambda/2$ on both particles $P_1$ and $P_2$. Although this effect is very weak, the phenomenon is more pronounced when the interparticle distance, $d$, is less than 10 µm. In this case, the backward scattered spherical waves from neighbouring particles propagate in the opposite direction to the incident light and interfere, causing short-period modulations of the optical forces. Since we neglect the backward scattered light when using the scattering-matrix method, the observed results are smooth curves with no short-period modulation of the scattering forces over long interparticle distances.

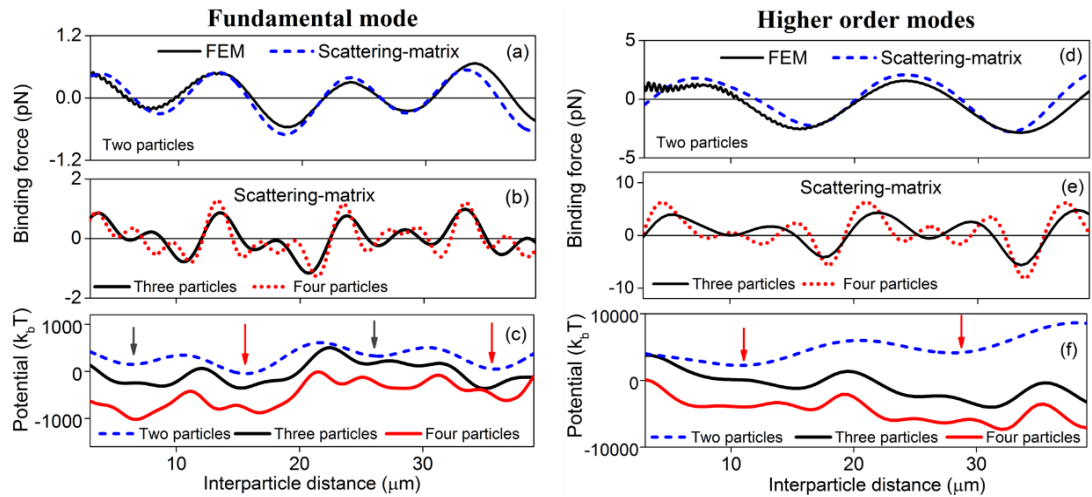

Figure 2. Optical binding forces and their corresponding potentials on two-, three- and four-

particle models. Binding forces of two particles in the evanescent fields of the FM (a) and HOMs (d) calculated from both the FEM and the scattering-matrix methods; (b & e) are the binding forces for three and four particles; (c & f) are the binding potentials of the corresponding binding forces. The potential is in units of $k_B T$, where $k_B$ is the Boltzmann's constant relating the thermal energy to an absolute temperature $T$ around the trapped particles. The fibre diameter is 2 μm.

Table 1 shows the parameters used for the scattering-matrix approach which gave us the best fit of the optical forces when compared to the FEM method. It is worth noting that here we assume that microparticles trapped in the evanescent fields of higher order microfibre modes scatter photons from one mode into another. In this case, each coupling coefficient, $\eta_i$, and the wave-vector component, $k_{tot}$, introduced in equations 2 & 3 represent different physical properties. Modifying these parameters will directly affect the fit of the binding forces, in both their shape and magnitude, for a given particle in a bounded particle chain at the fibre waist. We first choose the amplitudes of the incoming fields, $A_j$, to match the FEM force calculations for a chosen incident power of 30 mW. The wave-vector components, $k_b$ and $k_c$, are then used to fit the frequency oscillations, the peak positions and the damping parameters of the binding forces ($F_{n,j}$, $F_{b,j}$). The coupling coefficients, $\eta_4$, $\eta_5$ and $\eta_6$, of the scattering particles identify the curve shape and slopes of the mean force values and the propagation lossy modes between the trapped particles.

Table 1. Wave-vector length ratios of the scattered lights, $k_b$ and $k_c$, to the incident field, $k_a$ ($k_a = 2\pi/\lambda$, where $\lambda$ = 1064 nm). The coupling coefficients, $\eta_i$, are used in the scattering-matrix approach for both FM and HOM propagation in a 2 μm microfibre.

| | Wave-vector length ratio | | Coupling coefficients $\eta_i$ | | |
|---|---|---|---|---|---|
| Fundamental mode ($LP_{01}$) | $k_b/k_a$ | $k_c/k_a$ | $\eta_4$ | $\eta_5$ | $\eta_6$ |
| | 0.90 | 0.96 | 0.13 | 0.10 | 0.08 |
| Higher order modes ($LP_{11}$) | $k_b/k_a$ | $k_c/k_a$ | $\eta_4$ | $\eta_5$ | $\eta_6$ |
| | 0.93 | 0.86 | 0.11 | 0.16 | 0.09 |

In order to study the binding mechanisms for large particle ensembles on a microfibre surface, the scattering-matrix approach is applied. Here, we use the parameters of Table 1 to calculate the binding forces of three- and four-particle chain formations in the evanescent fields of a MNF system. The main assumption in these calculations is that the changes of the mode coupling strengths, $\eta_l$, and the wave-vector components, $k_b$ and $k_c$, among the scattering particles and the microfibre light modes are negligible when compared to the two-particle case. This is a reasonable assumption since the transmission loss of the incident beam due to the light scattering from each individual particle is negligible. If we additionally assume that the relative distances, *d,* between the next neighbouring particles within a chain are equal, we can also investigate the modulations of the binding force on each individual particle over the whole complex, bounded-particle chains using this very simple model. Figure 2 (b & e) show the variation of the binding forces for the end particle of a chain as a function of interparticle distances; for example, the third particle, $P_3$ (black curve), and the fourth particle, $P_4$ (red dotted curve), of the three- and four-particle chains. The binding forces on these end particles are greater and the oscillations are more frequent if the chain has more particles in it for both the FM and HOM cases.

Figure 2 (c & f) show the optical binding potentials calculated from the corresponding binding forces. Several potential well regions are created (indicated by the black and red arrows) where the particles could form multistable configurations due to these binding force modulations for both the FM and HOMs. For the two-particle case (dark blue dashed lines), the periods of the potential depth modulation are 10 µm and 16 µm for the FM and HOM fields, respectively. The larger stable potential distance of the HOMs compared to the FM is due to the longer extension of the HOM's electric field distribution in the surrounding medium. This first generic observation of the potential depths and their behaviours is similar to the previous reports using Bessel beams[5,8], as well as our earlier work using a nanofibre[30].

It is interesting to note that the FM potential plots for the two-particle case in Figure 2c (dark blue dashed line) exhibit shallower potential minima (~200 $k_BT$) at the positions of 6 µm and 26 µm (black arrows) when compared to the potential minima (~ 600 $k_BT$) at the positions of 16 µm and 37 µm (red arrows). These metastable positions (6 µm, 26 µm) only create long-range local oscillations of the particles before they eventually reach the deeper potential wells at 16 µm and 37 µm. In contrast, Figure 2f (dark blue dashed line) shows that the depths of the HOM's potential wells for two particles appear to be similar to one another, at approximately 2,500 $k_BT$, within the studied range of the interparticle distance. The expected stable interparticle distances for two particles are at 12 µm and 28 µm (red arrows), which are shorter than for the FM case. We can make initial conclusions here that a pair of particles trapped in the HOM evanescent fields in a microfibre system show (i) shorter stable interparticle distances, (ii) create deeper potential wells along the fibre waist, and hence, theoretically, (iii) exhibit stronger optical binding forces between the particles than compared to the FM propagating fields.

The black and red curves in Figure 2 (c & f) show the potential profiles of the bounded three- and four-particle chains. These potential wells exhibit complex patterns with varying potential minima depths over the large interparticle distance range along the fibre waist region for both the FM and HOMs cases. We see that the observed shallower potential wells strongly depend on the particle numbers within the chain and the chosen coupling strength, $\eta_6$, which describes the propagation loss outside the fibre. These shallow potential oscillations disappear when $\eta_6$ approaches zero.

The shallow potential wells, which cause the particles to oscillate, can be overcome by thermal activation. The particles would then approach the deeper potential regions. The stronger binding force observed due to additional particles would also cause the stable interparticle distances to fall until the new deepest potential well can be created. This results in closer localisations of the particles with slightly non-equilibrium interparticle distances in the chain. To confirm this hypothesis, we consider non-symmetrical conditions where the interparticle distances are not identical. For the three-particle chain calculations, Figure 3 (a & b) show the force lines of the particles, $P_2$ and $P_3$, as a function of the two interparticle distances, $d_{P1-P2}$ and $d_{P2-P3}$, for the case where the forces acting on all three particles are equal. Here, the distances $d_{P1-P2}$ and $d_{P2-P3}$ are assigned to the spacing between the particles, $P_1$ and $P_2$, and to the particles, $P_2$ and $P_3$, respectively. There are many intersections of these lines where the binding forces on particles vanish and the equilibrium configurations can be observed.

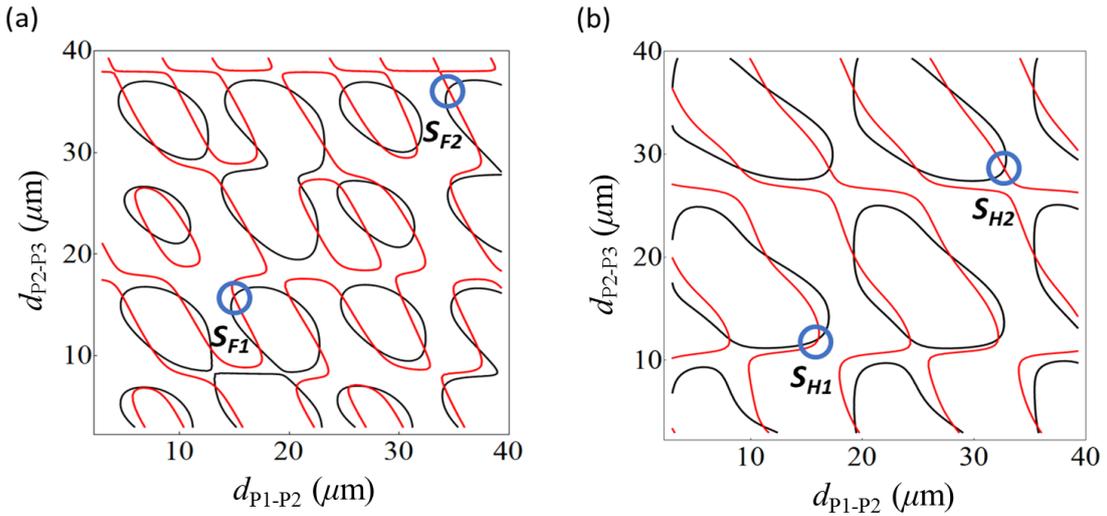

Figure 3. Contour plots for equilibrium positions of particles. Equivalent force lines for a three-particle chain formation as a function of the two independent interparticle distances in the FM (a) and HOM (b) evanescent fields. Intersections of these equal force contours indicate the equilibrium configurations but only those denoted by the blue circles are under stable conditions.

Although many equilibrium possibilities are denoted in Figure 3 (a & b), we consider that stable configurations of particles in the chains can only be achieved when they localise at a potential minimum deeper than the particles' thermal energy. In addition, the propelling particles along the fibre surface also increase the particle kinetic motions, and hence, would allow them to jump between the nodes before they approach the deeper potential minima. Assessing the results in Figure 2 (c & f) alongside Figure 3, we can finally predict that there are two major stable configurations which are considered the most preferable locations for the particles. As shown in Figure 3 (a & b), these two stable configurations (circled) correspond to the stable positions, $S_{F1}$ ($d_{P1-P2}$ = 15 µm, $d_{P2-P3}$ = 16 µm) and $S_{F2}$ ($d_{P1-P2}$ = 34.5 µm, $d_{P2-P3}$ = 36 µm) for the FM case, and the stable positions, $S_{H1}$ ($d_{P1-P2}$ = 16 µm, $d_{P2-P3}$ = 12 µm) and $S_{H2}$ ($d_{P1-P2}$ = 29 µm, $d_{P2-P3}$ = 32.5 µm) for the HOMs case. We notice that, although there are slight shifts of the observed $d_{P1-P2}$ compared to the $d_{P2-P3}$ values for the non-symmetrical configurations, they are still very close to the deep potential positions, as shown in Figure 2 (c & f) for the FM and HOMs.

**Experimental observation**

As shown in Figure 4, the experiment consists of three components: (i) *LG* beam generation and HOM excitation, (ii) tapered fibre fabrication, and (iii) the optical tweezers (see Methods). A specific number of particles was trapped using a time-sharing optical tweezer and all were brought close to the tapered fibre simultaneously. As soon as the particles are released from the tweezers trap, the fibre's evanescent field propels them along its axis.

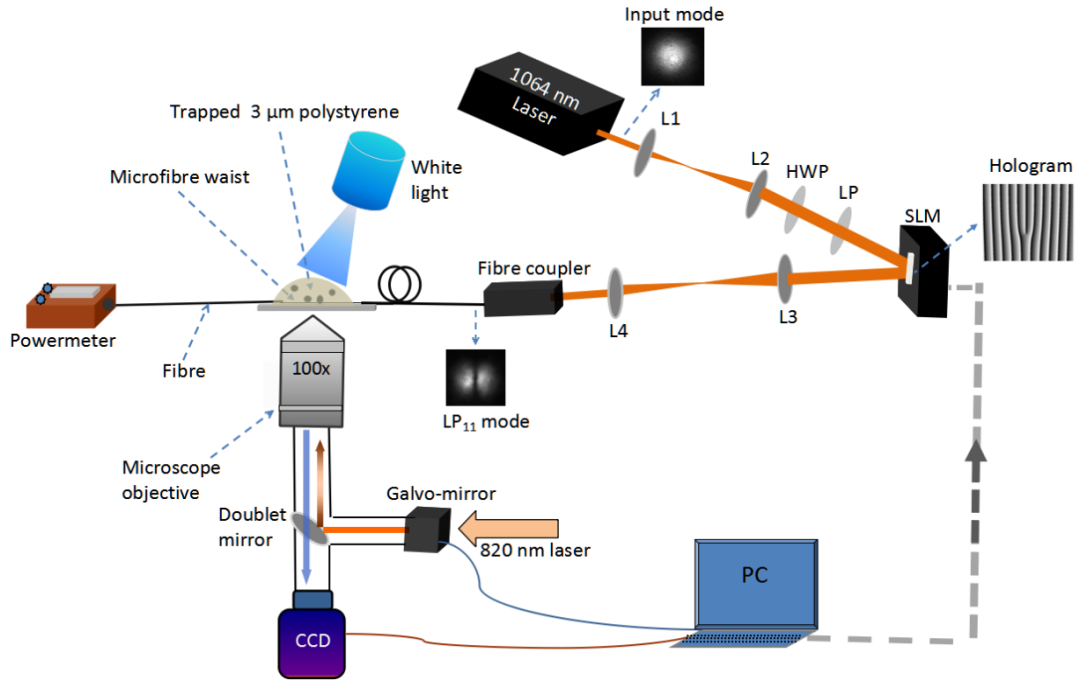

Figure 4. Experimental setup for particle propulsion. L1, L2, L3, L4: lenses; LP: linear polariser; HWP: half-wave plate; CCD: charge-coupled device camera; computer. Orange lines represent the free beam path.

Figure 5 is a micrograph of the particle speed under the influence of the propagating FM and HOM fields extracted from videos of particle motion (see Supplementary Movies S1, S2, S3 & S4 online as examples). Starting from the first particle, the incident laser beam interacts with all particles, labelled from $P_1$ to $P_5$. As is clearly seen in the micrographs, the particles self-arrange along the fibre. The more particles used, the smaller the interparticle distances between $P_1$, $P_2$ and $P_3$. Once the particles settled at their equilibrium positions, the interparticle distances for each particle pair were measured.

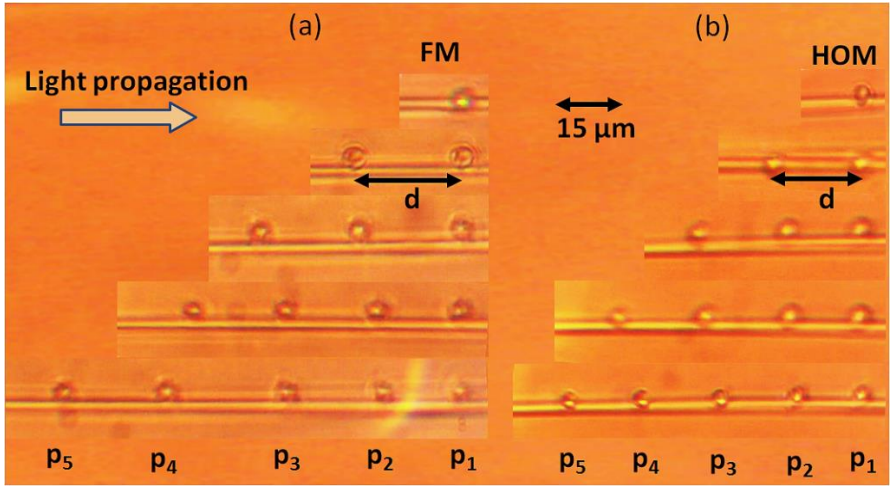

Figure 5. Micrograph of inter-particle distance for particle numbers changing from 1 to 5 in particle chains. (a) Fundamental mode propagation; (b) Higher order mode propagation. The power at the microfibre waist is $P_{in}$ = 30 mW.

Any small imperfection on the fibre surface may cause a local acceleration or deceleration of some particles in the chain. When this acceleration/deceleration breaks the stable interparticle distance, there is always an attractive or a repulsive kick on neighbouring particles to compensate the change in position and to return the system to equilibrium. This self-adjustment of the particle distance was far more obvious for the FM work than for the HOM studies. When we used HOMs, sometimes the particles were not able to re-establish their equidistance; as a result, they behaved as independent particles and left the chain.

As we discussed in the theoretical section, by assuming that the particles localise at the first stable configuration within the chains, and using the values derived in Figure 3, $d_{P1-P2}$, $d_{P2-P3}$, we plot in Figure 6a the calculated particle speeds (dashed curves) of up to four bounded particle chains using the standard bulk Stokes' drag coefficient, $F = 6\pi\mu a v$, where $\mu$ is the viscosity and $v$ is the particle velocity. The solid curves in Figure 6a show the experimental observation of up to five particle formation chains. We present in Figure 6b the theoretical and experimental data of the ratio between the corresponding speed of particles in the HOMs and FM fields. Taking a closer look at the particle speed ratios, the original speed ratio of 5.5 (experiment) and 5.4 (theory) for a single particle was found to have approximately halved to 3 (experiment) and 2 (theory) for four particles. Even though theory and experiment show some discrepancy, it is clear that they are somewhat in agreement.

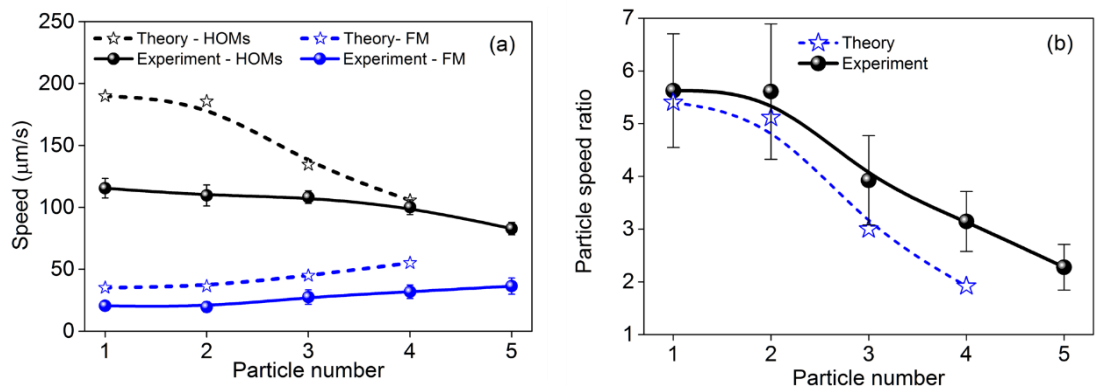

Figure 6. Speed comparison of particles under the FM and HOMs. (a) Theoretical data (dashed lines) and experimental observation (solid lines) of particle speed change with respect to the number of particles for both HOMs and FM. (b) Theoretical calculation (dashed lines) and experiment (solid lines) of particle speed ratio between the HOM and the FM cases.

In the case of FM propagation, changes to the particle speeds were slow, but the plot shows a clear trend toward higher speeds with increasing particle number. This is consistent with predictions in previous works[11,30]. However, this increase has a saturation limit, and the particle speed tends to be constant after five or six particles. Consequently, the binding between spheres was found to be weaker. Interestingly, in the case of HOM propagation, this speed trend was opposite to that observed for the FM propagation. The more particles present, the slower the observed particle speeds. The particles' kinetic motions sometimes allow them to escape from the stable configuration.

The interparticle distances between neighbouring particles in various lengths of particles chain are given in Figure 7 for the FM and HOMs. The experimentally obtained interparticle distance (solid curves) appear to be largest when only two particles are present in the chain and it decreases slowly as more particles are added. This self-adjustment of interparticle distance within the chain matches with the trend of the theoretical prediction (dashed curves). The discrepancy between the theoretical prediction and the experimental observation may be due to the fact that we calculate the optical forces on the particles using a 2D configuration for the original FEM method. When there are only two particles, they essentially share the incident beam with little scattering loss and the interparticle spacing can be set accordingly. As the particle number increases, the incident beam is distributed over the particles with certain ratios. The particles closest to the incident light source receive larger portions of the power than distant particles due to scattering losses. The non-uniform distribution of scattered light on the particles requires different interparticle distances in order to maintain the self-arranged chain of particles.

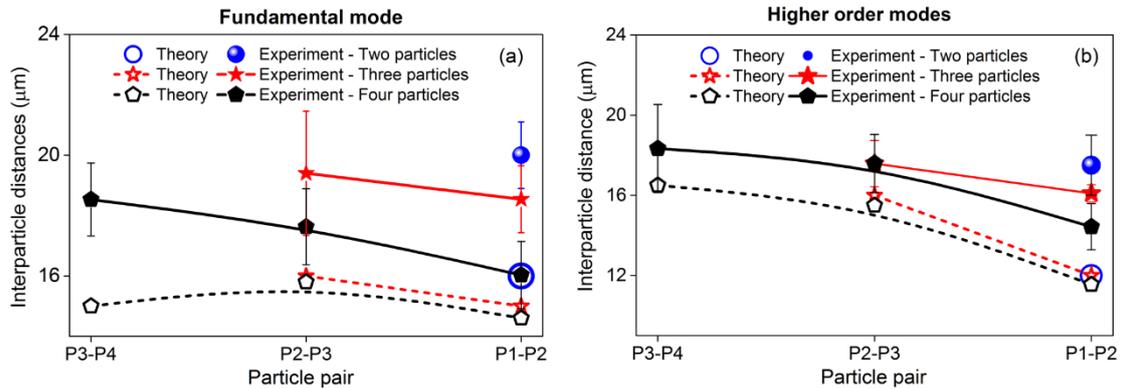

Figure 7. Interparticle distance with respect to the number of particles in a chain (in μm unit). Theoretical prediction (dashed lines) and experimental observation (solid lines) of interparticle distances between neighbouring particles in various length of particle chains under the FM (a) and HOMs (b) evanescent fields.

The experimental observation of interparticle distances from Figure 7 confirms our hypothesis that the particles position themselves around the first preferable stable position. An important factor to be gleaned from this result is that, regardless of the particle number, the interparticle distances are always slightly smaller when we use HOMs instead of the FM. This is in good agreement with the theoretical predictions. Smaller particle separations indicate that the stronger evanescent field intensity of the HOMs is not the important factor responsible for ordering the interparticle separation. Instead, the potential profiles and how the scattered light fields from the particles interfere must play a larger role in determining the interparticle separations.

An important distinction between the binding potentials of the FM and HOM mode cases can be realized from Figure 2. The interparticle stable positions are not only governed by the magnitude of the binding forces, but also the shape of the potential landscape. As shown in Figure 2 for the two-particle case, although the absolute binding potential is deeper under the HOMs when compared with the FM, this potential well is relatively wide, which may be responsible for uncertainties in the particles' positions. Additionally, when the number of particles within the chain increases, the potential landscape transforms to contain multiple potential wells with similar magnitudes rather than a single potential minimum, leading to particles jumping between these potential minima. This may explain why the particles under the HOMs exhibit relatively unstable interparticle separations.

Furthermore, when a dielectric particle is trapped in the evanescent field, the scattering force is responsible for the propulsion of particles along the fibre axis. To better understand how the binding force affects the control of the trapped particles in a chain, we calculate the ratios of the optical binding to the scattering forces. The ratio of the maximum absolute values of the binding force to the scattering force was found to be 0.7 for the FM and 0.5 for the HOMs. This smaller ratio could explain the case in which particles sometimes escape from a stable configuration in the HOM field. This implies that, although a stronger optical binding force is observed for the HOMs, it is still easier to control each individual particle within the particle chain using the scattering force. This is in contrast to the fundamental mode propagation.

In conclusion, we studied the optical binding effect for a number of 3 µm polystyrene particles under the influence of higher order mode propagation in an optical microfibre. By combining the FEM and a simple scattering-matrix approach, we were able to investigate the dynamics and the self-arrangements of particles in both the FM and HOM evanescent fields of MNF systems. In the FM case, the relatively larger interparticle distance and the rigid particle chains hint at a stronger interaction, which in turn, could enhance the speed. For HOM propagation, both theory and experimental results show a smaller interparticle distance and an instability of chains consisting of a large number (five) of particles. Comparing the observed behaviour with that obtained for FM propagation, a reasonable explanation of the particle speed and the interparticle distances can be provided. These interesting physical properties of HOMs offer a better understanding of the interactions of light with matter. The expected reduced optical interaction of HOMs due to the spin and the orbital angular momentum may make it a better candidate for 3D manipulation of micro and nano-objects. In particular, this study could be very useful for applications such as atom trapping and nanoparticle trapping, etc.

**Experimental Method**

**Higher order mode generation**

To create a beam with a doughnut-shaped intensity cross-section, a linearly polarised 1064 nm $Nd^{3+}$: YAG laser was launched onto the SLM (SLM-BNS 1064). A computer-generated vortex phase discontinuity combined with a blazed grating was applied to the SLM so that a first order Laguerre-

Gaussian ($LG_{01}$) beam was created in the far field. Two-mode fibre (Thorlabs, SM1250G80) operating at 1064 nm with a cladding diameter of 80 µm was chosen for the experiment. The fibre supports both the fundamental $LP_{01}$ and the $LP_{11}$ family of higher order modes. The $LG_{01}$ beam was coupled into the fibre and a two-lobed pattern corresponding to the $LP_{11}$ mode can be obtained at the fibre output. With the right objective lens, approximately 40% of the HOM power could be coupled into the fibre and a LabVIEW programme allowed us to easily switch between different orders of *LG* beams. By switching back to the FM, 70% of the output power was coupled into the fibre.

**Preparation of higher-mode tapered fibre**

A brushed hydrogen flame was used to make the tapered fibres, which can be customised to any desired taper shape[20,39,40]. A double linear taper with physical taper angles of 0.6 mrad and 1 mrad was predesigned and the fibre was fabricated with this profile. After pulling, ~80% transmission of the HOMs was achieved for a fibre with a 2 µm waist. The same fibre had 95% transmission for the FM mode. In real applications it is always more reasonable to state the power at the fibre waist. Assuming a symmetrical fibre taper, the power at the waist is estimated to be the square root of the product of the input and output powers[36].

**Integrating an optical tapered fibre into the optical tweezer**

The prepared fibre was mounted onto a U-shaped metal mount and attached to a 3D translational stage positioned in an optical tweezer. The taper's vertical and horizontal positions were adjusted over the trapping plane of the optical tweezer. The optical tweezer was also equipped with a galvo mirror array, controlled using a MATLAB code to achieve time sharing between the multiple traps. The reason for integrating the fibre into the optical tweezer system was to facilitate trapping of a specific number of particles and to minimise any disturbances due to unwanted particles[36]. A dilute 3 µm polystyrene particle dispersion was dropped onto the microfibre, which was located at the focal plane of the tweezer. First, the optical tweezer was used to trap targeted particles and to move them to the microfibre. Then the tweezer was switched off. The particles are attracted to the fibre by the evanescent field and are propelled along the waist region. By monitoring the particle motion via a camera (Thorlabs 1240), particle speeds and relative particle distances can be extracted. The same waist power (30 mW) was used for both the FM and the HOMs. The experiment was repeated three times for each set of particles in a chain, ranging from one to five. For each sequence, the speed and inter-particle distances of the corresponding particles in each chain were analysed for both the FM and the HOM evanescent fields.

**Acknowledgements**

This work was supported in part by the Okinawa Institute of Science and Technology Graduate University. H. R. and D. H. acknowledge the support of the Austrian Science Fund (FWF) through SFB Foqus Project F4013. This article is based upon work from COST Action MP1403 "Nanoscale Quantum Optics", supported by COST (European Cooperation in Science and Technology). The authors wish to thank M. Daly, P. S. Mekhail and S. D. Aird for useful comments about the manuscript.